\documentclass{elsart}
\usepackage{epsfig}

\begin{document}

\begin{frontmatter}

\title{Characterization of the anticipated synchronization regime in the coupled FitzHugh--Nagumo model for neurons}
\author{Ra\'ul Toral$^{1,3}$, C. Masoller$^{1,2}$, Claudio R. Mirasso$^{1}$,}  \author{M. Ciszak$^1$ and O. Calvo$^1$}
\address{$^1$ Departament de F{\'\i}sica, Universitat de les Illes Balears, E-07071 Palma de Mallorca, Spain }
\address{$^2$ Instituto de F{\'\i}sica, Facultad de Ciencias, 
Universidad de la Rep\'ublica, Igua 4225, Montevideo 11400, Uruguay}
\address{$^3$ Instituto Mediterr\'aneo de Estudios Avanzados, IMEDEA (CSIC-UIB), E-07071 Palma de Mallorca, Spain }

\begin{abstract}
We characterize numerically the regime of anticipated synchronization in the coupled FitzHugh-Nagumo model for neurons. We consider two neurons, coupled unidirectionally (in a master-slave configuration), subject to the same random external forcing and with a recurrent inhibitory delayed connection in the slave neuron. We show that the scheme leads to anticipated synchronization, a regime in which the slave neuron fires the same train of pulses as the master neuron, but earlier in time. We characterize the synchronization in the parameter space (coupling strength, anticipation time) and introduce several quantities to measure the degree of synchronization.
\end{abstract}

\begin{keyword}
Synchronization \sep excitability \sep noise \sep time delay 
\end{keyword}
\end{frontmatter}

Synchronization phenomena is a fascinating subject that has attracted a lot of attention in the last years\cite{PRK01}.
Recently, Voss\cite{V00,V01a,V01b,V02} has shown that it is possible to
synchronize autonomous dynamical systems in a master--slave configuration in
such a way that the slave system can actually anticipate (i.e. predict) the
trajectory of the master system. This result is surprising at first sight for
two facts: the dynamics of the master is not modified by the presence of the
slave, and the slave integrates its equations of motion at the same speed than
the master does. This remarkable phenomenon is achieved by the introduction of
appropriate delay lines in the dynamics of the slave system. More precisely, one of the schemes devised by Voss considers master ${\bf x}(t)$ and slave ${\bf
y}(t)$ (vector) dynamical systems, whose dynamics follow the general form:
\begin{equation}
\label{scheme1}
\begin{array}{rcl}
\dot {\bf x}(t)& =& {\bf f}({\bf x}(t))\\
\dot {\bf y}(t) & = & {\bf f}({\bf y}(t))+{\bf K}[{\bf x}(t)-{\bf y}(t-\tau)].
\end{array}
\end{equation}
The function ${\bf f}({\bf x})$ defines the dynamical system under consideration, ${\bf K}$ is the coupling strength matrix and $\tau$ is the delay time in the feedback loop of the slave system. As stated before, these equations admit the ``anticipated'' manifold \mbox{${\bf y}(t)={\bf x}(t+\tau)$} as a (structurally) stable solution\cite{V00}. This has been shown to be possible even in systems in which the dynamics, being chaotic, is highly unpredictable. Implementations of this result have been demonstrated theoretically \cite{M01} and experimentally \cite{LTDASL02} in unidirectionally coupled laser systems as well as in electronic circuits \cite{V02}. Some understanding of the anticipation mechanism can be achieved by the study of simple maps \cite{MZ01,HMM02}. 

\begin{figure}
\centerline{\makebox{\epsfig{figure=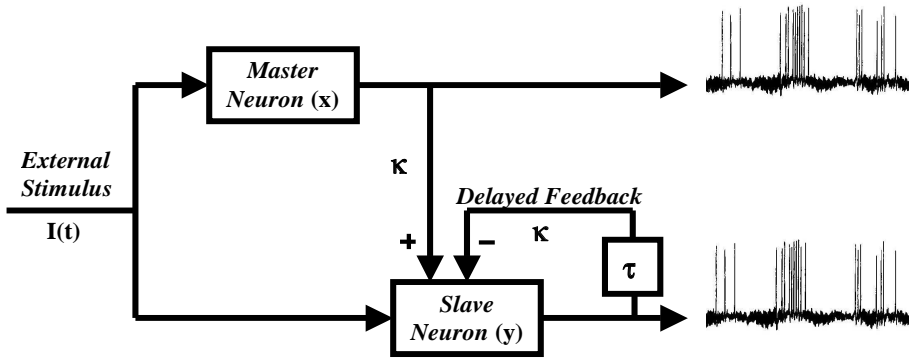,width=14cm,height=6cm,angle=0}}}\caption{\label{fig1}Schematic diagram of two model neurons coupled in a unidirectional configuration, subjected to the same external stimulus and with a feedback loop (with a delay time $\tau$) in the slave neuron.
}
\end{figure}

We have recently extended this result \cite{CCMMT02} by considering non--autonomous systems in which the dynamics is subjected to the effect of an external perturbation. Namely, we consider dynamical equations as:
\begin{equation}
\label{scheme2}
\begin{array}{rcl}
\dot {\bf x}(t)& =& {\bf f}({\bf x}(t))+{\bf I}(t)\\
\dot {\bf y}(t) & = & {\bf f}({\bf y}(t))+{\bf I}(t)+{\bf K}[{\bf x}(t)-{\bf y}(t-\tau)].
\end{array}
\end{equation}
where ${\bf I}(t)$ is an external input acting on both the master and slave systems (see scheme in Fig. \ref{fig1}).
Remarkably, although the manifold ${\bf y}(t)={\bf x}(t+\tau)$ is no longer an exact solution of the equations (\ref{scheme2}) (except in the case of a periodic input ${\bf I}(t)={\bf I}(t+\tau)$), it will be shown that the slave can actually predict some interesting part of the dynamics of the master. 

We have considered as our dynamical system ${\bf x}$ a simple model for neuronal response. Neurons are classical prototypes of excitable systems: their response to an external perturbation is highly non-linear and depends on its magnitude and timing. If the perturbation is small the system evolves back to the steady state; but if the perturbation exceeds a certain threshold, the system fires a pulse-like spike ({\it action potential}). Following the onset of the excitation, there is an interval during which another perturbation does not induce a new pulse ({\it refractory period}). Real neurons are complicated non-linear systems involving a large number of variables. Nevertheless, the essential features of their excitable behavior can be captured with a much-reduced description. The FitzHugh-Nagumo model provides the simplest representation of excitable firing dynamics and it has been widely used as a prototypic model\cite{K99,GHM91}.  

In this paper we study numerically the anticipated synchronization of two identical FitzHugh-Nagumo neurons, unidirectionally coupled, in the presence of a common external random forcing (see the schematic diagram shown in Fig. {\ref{fig1}). The model equations are:
\begin{equation}
\label{eq1}
\left\{\begin{array}{rcl}
\dot{x_1} & = & -x_1(x_1-a)(x_1-1)-x_2 + I(t)\\
\dot{x_2}& = & \epsilon (x_1-b x_2)
\end{array}\right.
\end{equation}
\begin{equation}
\label{eq2}
\left\{\begin{array}{rcl}
\dot{y_1} & = & -y_1(y_1-a)(y_1-1)-y_2 + I(t) + \kappa[x_1(t)-y_1(t-\tau)]\\
\dot{y_2}& = & \epsilon (y_1-b y_2) 
\end{array}\right.
\end{equation}
where {\bf x}=($x_1$,$x_2$) are the variables associated to the master neuron, {\bf y}=($y_1$,$y_2$) are the variables associated to the slave neuron, and $a$, $b$, and $\epsilon$ are constant parameters. $\kappa$
controls the strength of the coupling and $\tau$ is the delay time
associated to the feedback loop in the slave neuron. Note that only
the fast variables $x_1$, $y_1$ are coupled. The external common forcing $I(t)$ is a Gaussian random process of mean $I_0$ and delta--correlated in time (white noise): \makebox{$\langle [I(t)-I_0][I(t')-I_0]\rangle=D\delta(t-t')$}, $D$ being the noise intensity.

By choosing the mean value of the noise $I_0$ just below the threshold of the excitable system, a highly complex dynamics is observed. Spikes develop at random times in a completely unpredictable manner, see Fig. \ref{fig2}. The same figure shows that although the {\sl exact} details of the master dynamics are not reproduced by the slave, still it manages to anticipate the response of the master by firing its pulses just before the master does. 

\begin{figure}
\centerline{\makebox{\epsfig{figure=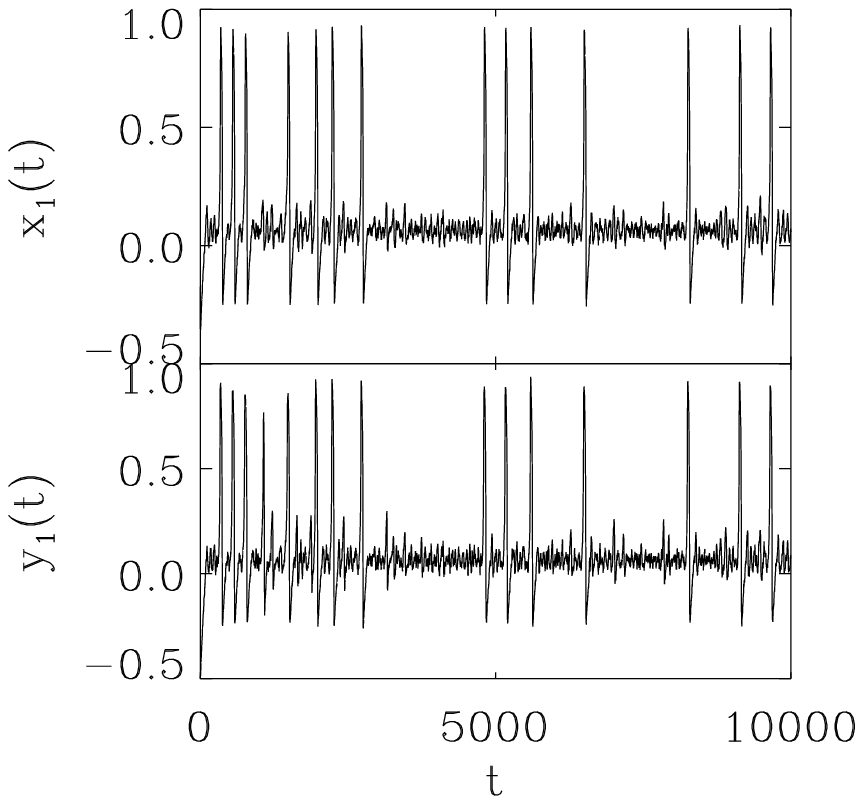,width=8cm,height=8cm,angle=0}\hspace{-1.0truecm}\epsfig{figure=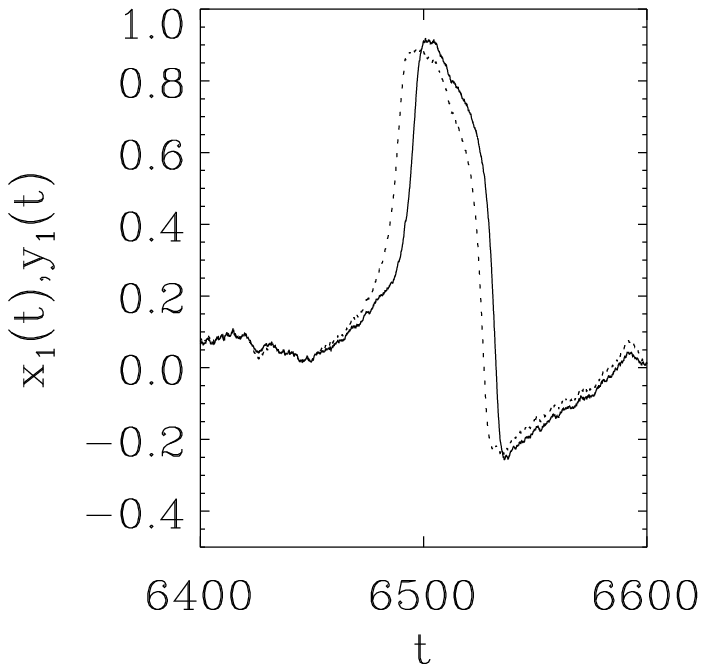,width=8cm,height=8cm,angle=0}}}
\caption{\label{fig2}Trains of spikes obtained from numerical simulations of Eqs. (\ref{eq1}-\ref{eq2}). The parameters are $a=0.139$, $b=2.54$, $\epsilon=0.008$, $I_0=0.03$, $\kappa=0.1$, $\tau=4$, $D=2.45\times 10^{-5}$. Left panel: Spikes of the master $x_1(t)$ and slave $y_1(t)$ neurons. Notice that the slave neuron makes an error around $t\sim 1000$ (4th pulse) in firing when the master does not. Right panel: detail of an anticipated spike. The solid line is the pulse of the master and the dotted line is the pulse of the slave.
}
\end{figure}

We now proceed to quantify the degree of synchronization. To this end we have measured the relative number of errors, $R$, made by the slave when anticipating the dynamics of the master. Notice in Fig. \ref{fig2} that the slave neuron occasionally fires an extra spike, which does not correspond to a spike fired by the master, but every spike fired by the master has a corresponding anticipated spike fired by the slave.  Thus, an error is defined as a pulse in the slave that has no corresponding pulse in the master.

\begin{figure}[!ht]
\centerline{\makebox{\epsfig{figure=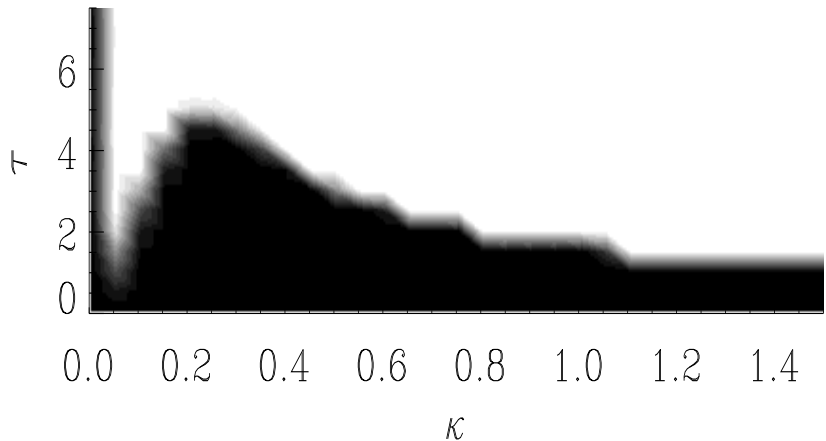,width=13cm,height=9cm,angle=0}}}
\vspace{-2.5truecm}\caption{\label{fig3} Relative number of errors $R$ in the parameter space ($\kappa,\tau$). The white region represents a region where the number of errors is larger than $R=0.1$.}
\end{figure}

Figure \ref{fig3} displays $R$ in a gray scale in the parameter space ($\kappa$, $\tau$). The dark (white) region represents a region where good (bad) synchronization occurs. In order not to miss too much detail, errors larger than $R=0.1$ have been uniformly plotted as white, while black indicates $R=0$, and the gray levels run between these two values. Two different synchronization mechanisms are present in Fig. \ref{fig2}. The first one appears for very low coupling intensity (the black region near the ordinate axis $\kappa=0$). This is not a regime of anticipated synchronization, but it corresponds to the synchronization of trajectories by common random forcing \cite{TMHP01} which leads simply to ${\bf x}(t)={\bf y}(t)$. 

Beyond this regime of synchronization by common random forcing, a finite value of the coupling $\kappa$ is required in order to achieve anticipated synchronization. However, a very large value of the coupling worsens the quality of the synchronization. This counter-intuitive result can be explained since the dynamics of the slave system becomes chaotic for such large values of the coupling and cannot follow the dynamics of the master. The existence of minimum and maximum values for the coupling in order to exhibit good anticipated synchronization agrees with what was previously found in autonomous chaotic systems \cite{V00} and in linear maps \cite{HMM02}. 

To quantify the anticipation time, we have computed the mean value $\langle t\rangle$ and standard deviation $\sigma$ of the time difference $t^m_i-t^s_i$, where $t^m_i$ are the times when the master neuron fires a pulse, and $t^s_i$ are the times when the slave neuron fires the corresponding pulse (hence the erroneous pulses fired by the slave are not taken into account). The data shown in the next figures are the result of averaging over a few thousand spike events.

Figure \ref{fig4}(a) plots the mean anticipation time $\langle t\rangle$ as a
function of $\tau$ for different values of the coupling $\kappa$. The results
for large $\kappa$ fall mainly on the line $\langle t\rangle=\tau$
corresponding to the anticipated solution \mbox{${\bf y}(t)={\bf x}(t+\tau)$}.
Notice that if $\kappa$ is small ('+' in Fig. \ref{fig4}(a)) it appears that $\langle t\rangle$ could even
be larger than $\tau$. However this result does not take into account that the
quality of the synchronization is poor in this case (it corresponds to the
grey region near the vertical axis in Fig. \ref{fig3}) and that the standard
deviation $\sigma$ is large (see Fig. \ref{fig5}) indicating a bad synchronization quality. Notice, finally, that for each
value of $\kappa$ there is a maximum anticipation time, in agreement with the
rather sharp transition between synchronized and desynchronized regimes shown in Fig. \ref{fig3}. 

Figure \ref{fig4}(b) plots the mean anticipation time $\langle t\rangle$ as a function of $\kappa$ for different values of $\tau$. The main result is that for each value of $\tau$ there is an interval of values of $\kappa$, $\kappa_{min}<\kappa<\kappa_{max}$, such that $\langle t\rangle \sim \tau$ (the plateaus in Fig. \ref{fig4}(b)). For small values of the coupling,  $\kappa<\kappa_{min}$ it is $\langle t\rangle \sim 0$, and this reflects that the two neurons are synchronized (not anticipately) due to the common external forcing (this parameter region corresponds to the dark region close to the vertical axis of Fig. \ref{fig3}). If $\kappa>\kappa_{max}$ the anticipation is lost due to the chaotic behavior of the slave. 

\begin{figure}[!ht]
\centerline{\makebox{\epsfig{figure=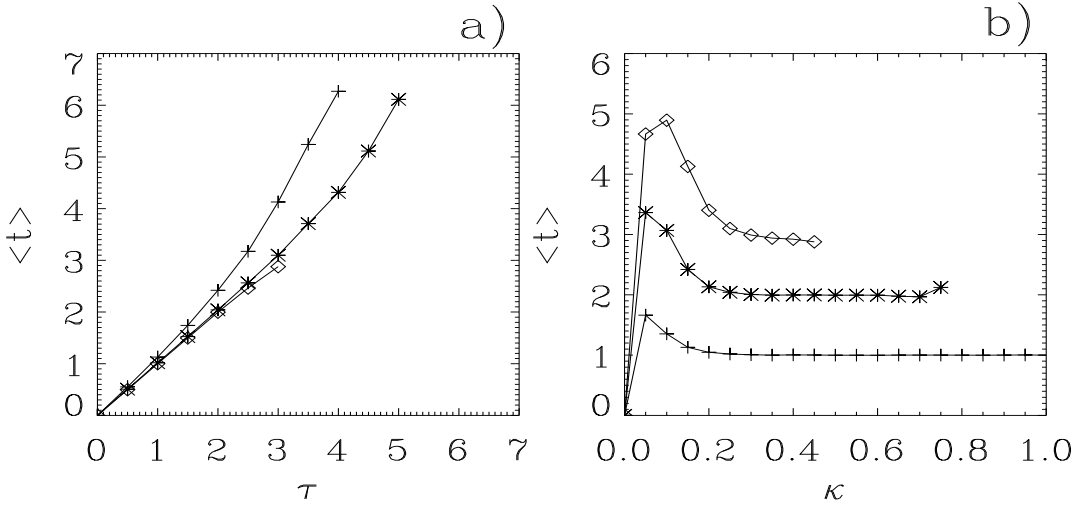,width=14cm,height=6cm,angle=0}}}
\vspace{0.5truecm}\caption{\label{fig4}(a) Mean anticipation time as a function of the delay time $\tau$ for the following values of the coupling strength: $\kappa=0.15$ (+), $\kappa=0.25$ (*) and $\kappa=0.45$ ($\Diamond$). (b) Mean anticipation time as a function of $\kappa$ for $\tau=1$ (+), $\tau=2$ (*) and $\tau=3$ ($\Diamond$). The results come from numerical integration of Eqs.(\ref{eq1}-\ref{eq2}) using the parameters indicated in Fig.\ref{fig2}.
}
\end{figure}

\begin{figure}[!ht]
\centerline{\makebox{\epsfig{figure=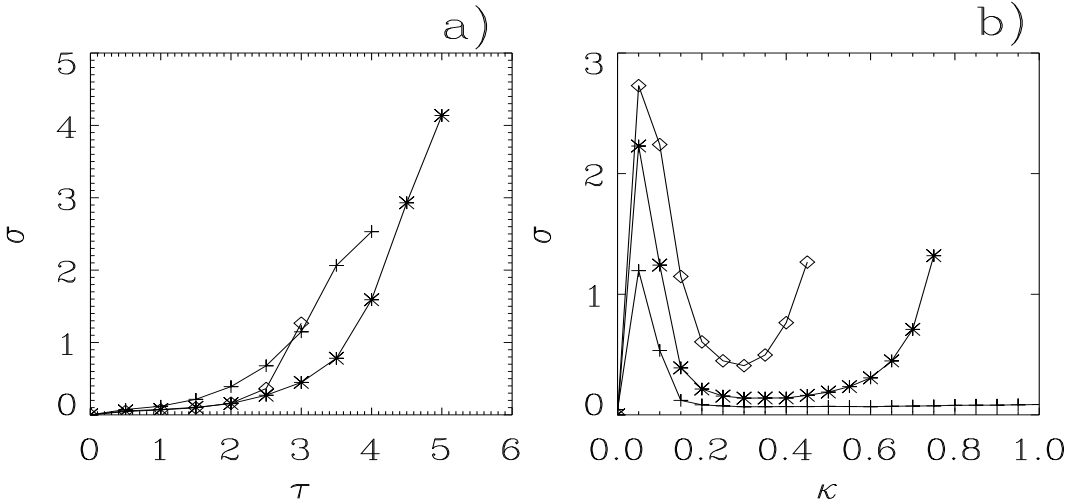,width=14cm,height=6cm,angle=0}}}
\vspace{0.5truecm}\caption{\label{fig5}Plot of the standard deviation of the anticipation time, $\sigma$, as a function of (a) the delay time $\tau$ and (b) the coupling $\kappa$ in the same cases as in Fig. \ref{fig4}.
}
\end{figure}

More information about the quality of the anticipated synchronization is obtained by looking at the dispersion in the values of $t^m_i-t^s_i$. In Figs. \ref{fig5}(a) and \ref{fig5}(b) we plot the standard deviation, $\sigma$, in the same cases as in Figs. \ref{fig4}(a) and \ref{fig4}(b). Of course, the best synchronization quality can be defined as the one with a small number of errors and a small dispersion in synchronization time. In this sense, one can see in Fig. \ref{fig5}(a) that $\sigma$ is an increasing function of $\tau$, indicating that the dispersion (and the quality of the synchronization) worsens for large $\tau$. Notice also in Fig. \ref{fig5}(b) that in the interval of coupling strength where good synchronization occurs, $\kappa_{min}<\kappa<\kappa_{max}$, $\sigma$ decreases significantly. 

In conclusion, we have studied numerically the regime of anticipated synchronization in coupled FitzHugh-Nagumo model neurons subjected to the same random external forcing. A difference with previous studies is that the anticipated synchronization manifold is not an exact solution of the dynamical equations. However, we have shown that the slave can predict the pulse firing quite accurately. We have introduced the normalized number of errors and the mean anticipation time to measure the degree of synchronization. We have shown that the anticipation phenomenon is robust and exists on a wide parameter region.

We would like to end by pointing out that synchronous neuronal oscillations underlie many cortical processes. Whether the results of this paper are of any interest to biological systems is open to speculation, but we hope that our numerical results will stimulate the search for anticipated synchronization in real biological neurons.

The work is supported by MCyT (Spain) and FEDER, projects BFM2001-0341-C02-01 and BMF2000-1108. C. Masoller acknowledges partial support from the Universitat de les Illes Balears.

\end{document}